\documentclass[prl,twocolumn,groupeness,showpacs]{revtex4}
\usepackage{amsfonts}
\usepackage{amsmath}
\usepackage{bm}
\usepackage{graphicx}
\begin{document}

\title{A Relation between $\theta_{13}$ and the Leptonic Dirac
CP Phase in $SO(10)$ Lopsided Models}

\author{Stephen M. Barr and Almas Khan}

\affiliation{Department of Physics and Astronomy, University of Delaware, Newark DE 19716}
\date{\today}
\begin{abstract}
It is shown that in $SO(10)$ models where the large solar and
atmospheric neutrino angles come from the charged-lepton mass matrix being
``lopsided", there is a characteristic relation between the 13
mixing angle of the neutrinos and the size of the Dirac CP-
violating phase in the lepton sector. This is illustrated in a
recently proposed realistic and predictive model.
\end{abstract}
 \maketitle

The symmmetries of the Standard Model do not constrain the masses
and mixings of the quarks and leptons, which are therefore free
parameters of that model. The best hope for obtaining predictions
(and testably precise postdictions) of these quantities seems to lie
with the more powerful symmetries of grand unified theories. The
greatest degree of predictivity comes from the unification group
$SO(10)$, since it relates the mass matrices of the up-type quarks,
down-type quarks, and charged leptons, and the Dirac mass matrix of
the neutrinos. (These four mass matrices will be denoted here by
$M_U$, $M_D$, $M_L$, and $M_N$, respectively.) An obstacle to making
predictions, however, is the fact that in the usual ``type I
see-saw" set-up \cite{seesaw} the mass matrix of the observed light
neutrinos $M_{\nu}$ depends on the Majorana mass matrix of the
right-handed neutrinos $M_R$ through the well-known see-saw formula
$M_{\nu} = - M_N M_R^{-1} M_N^T$. And because $M_R$ tends in most
models to be only loosely related by SO(10) symmetry, if at all, to
the other mass matrices, our experimental knowledge of the
properties of the quarks and charged leptons gives no information
about $M_R$. Since $M_R$ is unconstrained and is a symmetric
three-by-three complex matrix, it introduces many free parameters
into the calculation of the light neutrino masses and mixing angles.

In this letter, we discuss a simple class of $SO(10)$ models,
well-motivated on other grounds, in which the number of free
parameters coming from $M_R$ is much reduced and where it is
therefore possible to get definite predictions for the neutrino
mixing matrix $U_{MNS}$ \cite{mns}, including a testable relation
between two Standard Model quantities that have not yet been
measured, namely $\theta_{13}$ and $\delta_{lep}$. (We denote by
$\delta_{lep}$ the ``Dirac" CP-violating phase of the lepton sector.
For a review of leptonic CP violation, see \cite{leptoncp}.)

The neutrino mixing matrix is given by the standard formula $U_{MNS}
= U_L U_{\nu}^{\dag}$, where $U_L$ is the unitary transformation of
the left-handed charged leptons required to diagonalize the
charged-lepton mass matrix $M_L$, and $U_{\nu}$ is the unitary
matrix that diagonalizes the mass matrix of the observed light
neutrinos $M_{\nu}$. Typically, the matrix $U_L$ is highly
constrained or even known in $SO(10)$ models, because the unified
symmetry relates $M_L$ to the quark mass matrices; but in most
models $U_{\nu}$ is poorly constrained or unknown, because of its
dependence on $M_R$.  What is different about the models we are
discussing in this paper is that the neutrino Dirac mass
matrix $M_N$ is assumed to have negligibly small elements in its
first row and column (in the ``original basis" defined by the flavor
symmetries of the model). This obviously implies that $M_{\nu} = -
M_N M_R^{-1} M_N^T$ also has negligibly small first row and column,
which means that $U_{\nu}$ is in effect a $U(2)$ rather than a
$U(3)$ rotation. As such, it contains only one real rotation angle
and three complex phases, of which two phases do not contribute to
low-energy physics. In other words, in the kind of model we shall
discuss, only two parameters that depend on $M_R$ actually come into
the computation of the leptonic mixing matrix $U_{MNS}$, namely one
angle and one phase that we shall call $\theta_{\nu}$ and
$\phi_{\nu}$.

The crucial assumption that the first row and column of $M_N$ are
very small is motivated by two observations. First, in $SO(10)$
models there tends to be a close relationship between $M_N$ and the
mass matrix of the up-type quarks $M_U$. Second, in many models
$M_U$ has very small elements in its first row and column to account
for the extreme smallness of its smallest eigenvalue compared to its
largest: $m_u/m_t \sim 10^{-5}$, which is much less than the
corresponding ratios for $M_D$ and $M_L$: $m_d/m_b \sim 10^{-3}$ and
$m_e/m_{\tau} \sim 0.3 \times 10^{-3}$.

If, as we are assuming, $U_{\nu}$ is to a good approximation a
$U(2)$ rotation of the second and third families, then the large
solar neutrino angle, which involves the first family, must come
from the $U_L$ rather than from $U_{\nu}$. That is, the solar
neutrino angle must come from the diagonalization of $M_L$ rather
than $M_{\nu}$. But this means that in the original basis $M_L$ has
large off-diagonal elements, which is the distinguishing feature of
so-called ``lopsided models" \cite{bb1996, abb, lopsided2, bpw}. In
particular, one is naturally led to models of the ``doubly lopsided"
form \cite{bb1996, doublylop2, barr2007, barr2008, bk}.

The basic idea of ``lopsided models" is that large neutrino mixing
angles are caused by large asymmetrical off-diagonal elements in
$M_L$. All lopsided models explain the large atmospheric angle by
the 23 element of $M_L$ being large, i.e. as large as the 33
element. In doubly lopsided models the 13 element of $M_L$ is also
assumed to be large to explain the large solar angle. If these large
elements arise in an $SU(5)$-invariant way (i.e. from effective
operators of the form $C_1 {\bf 10}_1 \overline{{\bf 5}}_3 \langle
\overline{{\bf 5}}_H \rangle + C_2 {\bf 10}_2 \overline{{\bf 5}}_3
\langle \overline{{\bf 5}}_H \rangle + C_3 {\bf 10}_3 \overline{{\bf
5}}_3 \langle \overline{{\bf 5}}_H \rangle$), then the matrices
$M_L$ and $M_D$ have the form

\vspace{-0.25cm}
\begin{equation}
\begin{array}{l}
M_L = \left( \begin{array}{ccc}
- &  - & C_1 \\ - & - & C_2 \\ - & - & C_3 \end{array} \right) v_d ,\ \
M_D = \left( \begin{array}{ccc}
- & - & - \\ - & - & - \\ C_1 & C_2 & C_3 \end{array} \right) v_d
\end{array}
\end{equation}

\noindent where the dashes indicate elements much smaller than the
$C_i$. (The convention we use is that the left-handed fermions
multiply the mass matrix from the left, and the right-handed
fermions multiply it from the right.) The forms in Eq. (1) reflect
the well-known fact that $SU(5)$ relates $M_L$ to the transpose of
$M_D$. The reason for this left-right transposition is that the
${\bf 10}$'s of $SU(5)$ contain the left-handed down-type quarks and
right-handed charged leptons, while the $\overline{{\bf 5}}$'s
contain the right-handed down-type quarks and left-handed charged
leptons. That is why the large lopsided mass-matrix elements
$C_{1,2}$ produce large mixing of the {\it left}-handed leptons but
of the {\it right}-handed quarks, which accounts for the fact that
the MNS angles are big and the CKM angles are small.

The large elements of $M_L$ can be diagonalized by two successive
rotations of the left-handed charged leptons:

\vspace{-0.2cm}
\begin{equation}
\begin{array}{ll}
\left( \begin{array}{ccc} - & - & C_1 \\
- & - & C_2 \\ - & - & C_3 \end{array} \right) & \begin{array}{c}
U_{12}(\theta_s) \\ \longrightarrow \\ \end{array} \left(
\begin{array}{ccc} - & - & 0 \\ - & - & \sqrt{|C_1|^2 + |C_2|^2} \\
- & - & C_3 \end{array} \right) \\
 &  \begin{array}{c}\\ U_{23}(\theta_a)  \\
\longrightarrow \\
\end{array} \left(
\begin{array}{ccc} - & - & 0 \\ - & - & 0 \\
- & - & C \end{array} \right)
\end{array}
\end{equation}



\noindent where $C \equiv \sqrt{|C_1|^2 + |C_2|^2 + |C_3|^2}$, $\tan
\theta_s = C_1/C_2$ and $\tan \theta_a = \sqrt{C_1^2 + C_2^2}/C_3$.
Another rotation of the left-handed charged leptons (call it
$U'_{12}(\eta)$) is required to eliminate the small 12 element that
remains after the first two rotations. (The small elements that
remain {\it below} the main diagonal are eliminated by rotations of
the {\it right-handed} leptons.) Thus $U_L$ has the form $U_L =
U'_{12}(\eta) U_{23}(\theta_a) U_{12}(\theta_s)$.

The magnitude of the third rotation angle, $\eta$, depends on the
relative magnitudes of the small elements of $M_L$ that are denoted
by dashes in Eq. (1). If, as in the models we shall be considering,
the 32 element of $M_L$ is much larger than the 22 and 12 elements,
then the angle $\eta$ is small, and $U_L$ has the approximate form

\vspace{-0.1cm}
\begin{eqnarray}
U_L & \cong & \left( \begin{array}{ccc} 1 & 0 & 0 \\
0 & c_a & s_a \\ 0 & -s_a & c_a \end{array} \right)
\left( \begin{array}{ccc} c_s & s_s & 0 \\ -s_s & c_s & 0 \\
0 & 0 & 1 \end{array} \right) \nonumber \\
& = & \left( \begin{array}{ccc} c_s & s_s
& 0 \\ - c_a s_s & c_a c_s & s_a \\ s_a s_s & -s_a c_s & c_a
\end{array} \right)
\end{eqnarray}

\noindent where $s_a \equiv \sin \theta_a$, $c_a \equiv \cos
\theta_a$, $s_s \equiv \sin \theta_s$, and $c_s \equiv \cos
\theta_s$. If the angles in $U_{\nu}$ are small (as they will be in
the models we are considering, because of the hierarchical nature of
$M_{\nu}$), then $U_{MNS} = U_L U^{\dag}_{\nu}$ will be
approximately given by Eq. (3). One sees from this that the doubly
lopsided structure accounts in a simple and natural way for the
``bi-large" form of $U_{MNS}$, i.e. the form in which the solar and
atmospheric angles are large, but the 13 mixing, $U_{e3}$ is small.
Note, however, that the angles $\theta_a$ and $\theta_s$ in Eq. (3)
are not exactly equal to the atmospheric and solar neutrino angles,
which we will denote by $\theta_{atm}$ and $\theta_{sol}$, since the
latter get small contributions from $\eta$ and from the angles in
$U_{\nu}$.

In the models we are discussing, $U_L$, can be determined from the
known quark masses, charged lepton masses and the CKM angles. That
means that $U_{MNS}$ depends only on the two unknown parameters
$\theta_{\nu}$ and $\phi_{\nu}$ coming from $U_{\nu}$. Since the
solar neutrino angle $\theta_{sol}$ tends to be quite insensitive to
these parameters, as will be seen, one has three observable
quantities in $U_{MNS}$, ($\theta_{atm}$, $\theta_{13}$, and
$\delta_{lep}$) being calculable in terms of just two free
parameters, thus yielding one prediction, which can be expressed as
a relation between $\theta_{13}$ and $\delta_{lep}$.

To see what kind of relation one expects, let us neglect $\eta$ and
approximate $U_L$ by the simple form in Eq. (3). Then

\vspace{-0.5cm}
\begin{equation}
U_{MNS}  \cong  \left( \begin{array}{ccc} c_s & s_s & 0 \\ - c_a s_s
& c_a c_s & s_a
\\ s_a s_s & -s_a c_s & c_a
\end{array} \right) \left( \begin{array}{ccc} 1 & 0 & 0 \\
0 & c_{\nu} &  s_{\nu} e^{i \phi_{\nu}} \\
0 & -s_{\nu} e^{- i \phi_{\nu}} & c_{\nu}
\end{array} \right).
\end{equation}

\noindent Here, $s_{\nu} \equiv \sin \theta_{\nu}$ and $c_{\nu}
\equiv \cos \theta_{\nu}$. Multiplying this out, one obtains (for
small $\theta_{\nu}$):

\begin{equation}
\begin{array}{cl}
\sin \theta_{sol} & \cong c_{\nu} s_s \cong \sin \theta_s, \\
\sin \theta_{atm} & \cong |c_{\nu} s_a + s_{\nu} c_a c_s e^{i
\phi_{\nu}}|
\\ & \cong \sin \theta_a + \cos \theta_a \cos \theta_{sol}
\sin \theta_{\nu}
\cos \phi_{\nu}, \\
\sin \theta_{13} & \cong s_s s_{\nu} \cong \sin \theta_{sol} \sin
\theta_{\nu},
\\ \delta_{lep} & \cong \phi_{\nu}.
\end{array}
\end{equation}

\noindent The last of these equations results from the fact that the
13 element of $U_{MNS}$ has a phase $\phi_{\nu}$, whereas the phases
of the other elements appear in terms that are subleading in the
small quantity $\sin \theta_{\nu}$. We may rewrite the second
equation in Eq. (5) as

\begin{equation}
\sin \theta_{\nu} \cong \frac{\Delta}{\cos \phi_{\nu}}, \;\;\;
\Delta \equiv \frac{\sin \theta_{atm} - \sin \theta_a}{\cos \theta_a
\cos \theta_{sol}}.
\end{equation}

\noindent Therefore

\vspace{-0.7cm}
\begin{equation}
\sin \theta_{13} \cong \frac{\sin \theta_{sol} \Delta}{\cos
\delta_{lep}}.
\end{equation}

In realistic doubly lopsided models based on $SO(10)$ \cite{abb,
bk}, it is typically found by fitting the quark masses and mixing
angles that $\theta_a \sim \pi/3$, so that $\Delta \sim 0.25$. A
graph of Eq. (7) using this value is shown in Fig. 1.
\begin{figure}
\includegraphics[scale=.7]{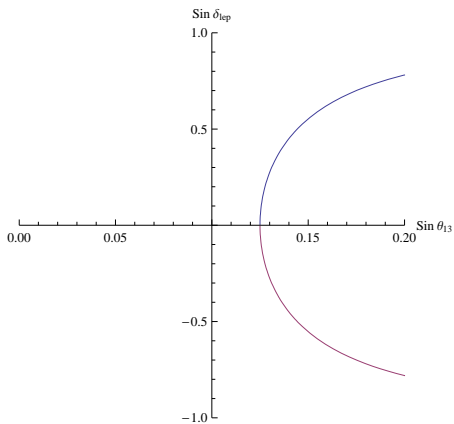}
\caption{The relation between $\sin \delta_{lep}$ and $\sin \theta_{13}$
 given in Eq. 7 for $\Delta = 0.25$.}
\label{{Fig.1}}
\end{figure}

We now illustrate these ideas in a particular model that is both
predictive and realistic model \cite{barr2007, bk}. It is a
non-supersymmetric $SO(10)$ grand unified model, in which the mass
hierarchy among the families arises radiatively; i.e. the masses of
the second and third families arise at tree level and the masses of
the first family from loop diagrams. The details of this model are
set forth in other papers \cite{barr2007, bk}; here, we merely
summarize. The mass matrices in this model have the approximate form
(we use slightly different notation than in \cite{bk}):

\begin{equation}
\begin{array}{ll}
\frac{M_U}{v_u} = \left( \begin{array}{ccc} 0 & 0 & 0 \\
0 & 0 & \frac{\epsilon}{3} \\ 0 & - \frac{\epsilon}{3} & 1
\end{array} \right),  & \frac{M_D}{v_d} = \left( \begin{array}{ccc}
0 & 0 & \delta \\ 0 & \delta_H & \frac{\epsilon}{3} + \delta'  \\
f C_1 & f C_2 - \frac{\epsilon}{3} & 1 \end{array} \right) \\ & \\
\frac{M_N}{v_u} = \left( \begin{array}{ccc} 0 & 0 & 0 \\
0 & 0 & - \epsilon \\ 0 & \epsilon & 1 \end{array} \right), &
\frac{M_L}{v_d} = \left( \begin{array}{ccc}
0 & 0 & C_1 \\ 0 & f_H \delta_H & C_2 - \epsilon \\
3 \delta & \epsilon + 3 \delta' & 1
\end{array} \right)
\end{array}
\end{equation}

\noindent where $\delta' \equiv (C_2/C_1) \delta$. The parameters
denoted by $\epsilon$, $\delta$, and $\delta_H$ are small, so that
$M_L$ and $M_D$ have the forms given in Eq. (1) (with $v_d$ and
$v_u$ scaled to make $C_3 =1$), and $M_U$ and $M_N$ indeed have
approximately vanishing first row and column.

The elements in these matrices denoted by $1$, $\epsilon$, and
$C_i$, $i=1,2$ arise at tree-level from three effective operators:
$O_1 = {\bf 16}_3 {\bf 16}_3 {\bf 10}_H$, $O_2 = {\bf 16}_2 {\bf
16}_3 {\bf 10}_H {\bf 45}_H/M_2$, and $O_3 = c_i {\bf 16}_i {\bf
16}_3  {\bf 16}_{iH}{\bf 16}'_H/M_3$ (i=1,2), respectively. Some of
the structure of these tree-level elements is easily understood
group-theoretically. The vacuum expectation value (VEV) of the
adjoint Higgs field $\langle {\bf 45}_H \rangle$ in $O_2$ is
proportional to the $SO(10)$ generator $B-L$ and gives the factor of
$-\frac{1}{3}$ in the $\epsilon$ terms of the quark matrices
relative to the lepton matrices. That factor is responsible for the
well-known Georgi-Jarlskog relation of quark to lepton masses
\cite{gj}. In $O_3$, the fact that the Higgs fields are in spinors
(${\bf 16}$) of $SO(10)$, which contain $\overline{{\bf 5}}$ but not
${\bf 5}$ of $SU(5)$, explains why this operator contributes the
elements $C_i$ only to $M_D$ and $M_L$, and not to $M_U$ or $M_N$.

The elements denoted by $\delta$, $\delta'$, and $\delta_H$ arise
from one-loop diagrams. $f$ and $f_H$ are factors reflecting the
breaking of $SO(10)$ (or more precisely, of $SU(4)_c$). The
parameters $v_d$ and $v_u$ set the overall scales of the mass
matrices of the $I_3 = -\frac{1}{2}$ and $I_3 = + \frac{1}{2}$
fermions, and have a small ratio ($v_d/v_u \simeq 0.9 \times
10^{-2}$) that is responsible for the small ratio of $m_b$ to $m_t$.
($v_d$ and $v_u$ come respectively from the VEVs of the $SU(5)$
$\overline{{\bf 5}}$ and ${\bf 5}$ in the $SO(10)$ ${\bf 10}$ of
Higgs fields.) Aside from this ratio of VEVs, all dimensionless
parameters of the model that come into the quark and lepton mass
ratios and mixing angles are of order 1 if they arise at tree-level,
and of order $1/16 \pi^2$ if they arise at one-loop level: a fit of
the data \cite{bk} gives $\epsilon \cong 0.189$, $C_1 \cong 1.03$,
$C_2 \cong -1.51$, $f \cong 0.566$, $f_H \cong 0.208$, $\delta \cong
2.29 (16 \pi^2)^{-1}$, and $\delta_H \cong 2.66 (16 \pi^2)^{-1}$.
Some of these parameters are complex. There are four physical
phases, but of these only two have a significant effect on the fit,
and these also are of order 1: $\arg(\epsilon) \cong 1.52$ rad and
$\arg(\delta_H) \cong 0.514$ rad.

In spite of the fact that the dimensionless parameters of the model
have ``natural" values, there is a good fit with 11 parameters to 14
measured quantities that span a very wide range: namely the quark
masses, charged lepton masses, CKM parameters, and the solar and
atmospheric angles. Of course, from the four Dirac mass matrices in
Eq. (8), it is not the angles in $U_{MNS}$ that are predicted, but
the angles in $U_L$. From the fit to the data performed in
\cite{bk}, the best fit value of $(U_L)_{23}$ comes out to be
$0.891$, whereas the experimental central value of $(U_{MNS})_{23}
\equiv \sin \theta_{atm}$ is about 0.71.

Since the matrices $M_U$ and $M_N$ in Eq. (8) have vanishing first
row and column, the mass of the up quark is zero at this level. The
up quark mass can be fit by a 11 element of $M_U/v_u$ that is of
order $10^{-5}$. That is too small to be a one-loop effect, but it
is the right magnitude to be a two-loop or three-loop effect. One
expects from $SO(10)$ symmetry that in $M_N/v_u$ there would also be
a 11 element of order $10^{-5}$. That should have negligible effect
on the predictions for the neutrino mixing parameters that will be
presented below.

To obtain predictions for the angles $\theta_{13}$ and
$\delta_{lep}$, we fix the parameters appearing in Eq. (8) to the
values that give the best $\chi^2$ fit to the following set of
measured parameters: quark masses, CKM angles, CKM phase, charged
lepton masses, and solar neutrino angle. This numerical fit was done
in \cite{bk}, and the details can be found in that paper. We then
scan over the possible values of $\theta_{\nu}$ and $\phi_{\nu}$ for
those that give a particular value of the atmospheric neutrino angle
$\theta_{atm}$ and plot the resulting points in the
$\theta_{13}$-$\delta_{lep}$ plane. This is shown in Fig. 2, where
the best-fit points coalesce to form the dark curves in the center
of the shaded bands. These curves, as expected, are similar to the
one shown in Fig. 1. Each shaded band represents a different value
of $\sin^2 \theta_{atm}$. The central curve in each band corresponds
to the parameters that give the best $\chi^2$ for the set of
measured parameters, which is $4.5$. The shaded region corresponds
to fits for which $\chi^2 \leq 6.5$.
\begin{figure}
\includegraphics[scale=.7]{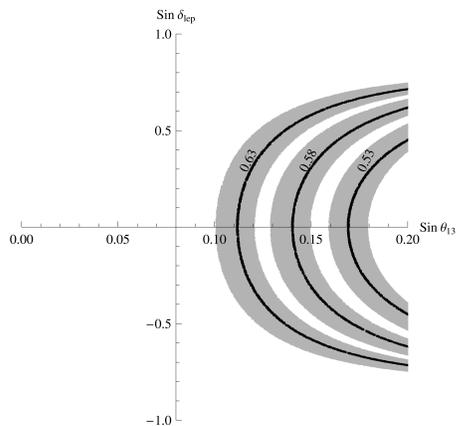}
\caption{The actual relation between $\sin \delta_{lep}$ and
 $\sin \theta_{13}$ for various values of $\sin^2 \theta_{atm}$
 (the numbers in the shaded band) in a realistic model}
\label{{Fig 2}}
\end{figure}

The major source of the uncertainty shown by the shaded bands is the
phase of the parameter $f_H$ in Eq. (8). This phase has no direct
effect on the quark masses and mixing angles and a relatively weak
effect on most of the leptonic quantities due to the fact that $f_H$
is such a small parameter. (Of course, $\arg f_H$ does have an {\it
indirect} effect on the quark masses and mixing angles, since the
effects on the very precisely known charged lepton masses from
varying $\arg f_H$ have to be compensated in the $\chi^2$ fit by
changes in the other parameters.) Since $\arg f_H$ can vary
considerably without harming the $\chi^2$ fit to the other
quantities, it has a significant effect on $\theta_{13}$ and
$\delta_{lep}$.

In conclusion, doubly lopsided models based on $SO(10)$ in which the
Dirac neutrino mass matrix has very small first row and column can
give interesting and testable predictions for the two as-yet-unknown
parameters of the Standard Model, $\theta_{13}$ and $\delta_{lep}$.
In particular, there is a fairly precise relation between these two
quantities, such that the smaller $\theta_{13}$ is the smaller is
$\delta_{lep}$, with a lower limit on $\theta_{13}$, as Figs. 1 and
2 show. These features have been illustrated in a particular
realistic model, which is non-supersymmetric and has a radiative
fermion mass hierarchy. However, qualitatively similar predictions
should also be obtainable from doubly lopsided $SO(10)$ models that
are supersymmetric and that have tree-level hierarchies. The
predicted relation between $\theta_{13}$ and $\delta_{lep}$ will
become more precise as the quark masses, CKM angles, CKM phase, and
solar and atmospheric neutrino angles are determined with more
precision. If they were known perfectly, the predicted relation
would be a single sharp curve like the one shown in Fig. 1. One
sees, then, that the rigorous testing of models of quark and lepton
masses will require progress along a broad front.

We acknowledge useful conversations with Ilja Dorsner. This research was supported by the DOE grant DE-FG02-
91ER40626 and partially by Bartol Research Institute.

\end{document}